\documentclass[pra,twocolumn,superscriptaddress,longbibliography]{revtex4-1}
\usepackage{amssymb,amsmath,amsfonts,bm}
\usepackage{graphicx}
\usepackage{epstopdf}
\usepackage{times}
\usepackage{float}
\usepackage{lipsum}
\usepackage{color}
\usepackage{ulem}

\usepackage{hyperref}
\hypersetup{colorlinks=true, linkcolor=blue, citecolor=red, urlcolor=blue  }

\usepackage{physics}

\begin{document}

\title{The Critical Behavior of Quantum Stirling Heat Engine}% 

\author{Yuan-Sheng Wang}
 \affiliation{School of Systems Science, Beijing Normal University, Beijing 100875, China}
 \affiliation{Lanzhou Center for Theoretic Physics, Lanzhou University, Lanzhou, Gansu 730000, China}
 
\author{Man-Hong Yung}  %\email{yung@sustech.edu.cn}
\affiliation{Department of Physics, Southern University of Science and Technology, Shenzhen 518055, China}
\affiliation{Shenzhen Institute for Quantum Science and Engineering, Southern University of Science and Technology, Shenzhen 518055, China}
\affiliation{Guangdong Provincial Key Laboratory of Quantum Science and Engineering, Southern University of Science and Technology, Shenzhen 518055, China}
\affiliation{Shenzhen Key Laboratory of Quantum Science and Engineering, Southern University of Science and Technology, Shenzhen,518055, China}

\author{Dazhi Xu} \email{dzxu@bit.edu.cn}
 \affiliation{Department of Physics, Beijing Institute of Technology, Beijing 100081, China}

\author{Maoxin Liu} \email{mxliu@bnu.edu.cn}
 \affiliation{School of Systems Science, Beijing Normal University, Beijing 100875, China}
 %\affiliation{Department of Physics, Department of Physics, Beijing University of Posts and Telecommunications, Beijing 100876, China}
 
\author{Xiaosong Chen} \email{chenxs@bnu.edu.cn}
\affiliation{School of Systems Science, Beijing Normal University, Beijing 100875, China}

\date{\today}% It is always \today, today,

\begin{abstract}
We investigate the performance of a Stirling cycle with a working substance (WS) modeled as the quantum Rabi model (QRM), exploring the impact of criticality on its efficiency. Our findings indicate that the criticality of the QRM has a positive effect on improving the efficiency of the Stirling cycle. Furthermore, we observe that the Carnot efficiency is asymptotically achievable as the WS parameter approaches the critical point, even when both the temperatures of the cold and hot reservoirs are finite. Additionally, we derive the critical behavior for the efficiency of the Stirling cycle, demonstrating how the efficiency asymptotically approaches the Carnot efficiency as the WS parameter approaches the critical point. Our work deepens the understanding of the impact of criticality on the performance of a Stirling heat engine.
\end{abstract}

                                  
\maketitle

%\tableofcontents

%break was forced \lowercase{via} \textbackslash\textbackslash}
\section{Introduction}
Recently, there has been a growing interest in exploring the application of thermodynamics in the quantum regime, thanks to advancements in experimental control over various quantum systems \cite{gemmer2009quantum, binder2018thermodynamics}.
While classical thermodynamics traditionally focused on large systems governed by classical physics, the emergence of quantum heat engines (QHEs) has provided a valuable platform for testing the principles of thermodynamics in the quantum realm. 
QHEs operate by utilizing quantum effects in either the reservoir or the working substance (WS) to convert heat into work through a thermodynamic cycle.
Extensive research has been conducted in the field of QHEs \cite{PhysRevE.72.056110,PhysRevE.76.031105,GELBWASERKLIMOVSKY2015329,doi:10.1116/5.0083192,cangemi2023quantum},
demonstrating that quantum effects, such as quantum coherence \cite{Scully15097,PhysRevX.7.031044,PhysRevLett.122.110601,PhysRevLett.127.190604,PhysRevApplied.6.024004,Korzekwa_2016,Kammerlander2016}, quantum correlation \cite{PhysRevX.5.041011,PhysRevLett.128.090602} and energy quantization \cite{PhysRevLett.120.170601}, can be harnessed to enhance their performance.

QHEs have been implemented in various experimental platforms,
such as cold atoms \cite{doi:10.1126/science.1242308,PhysRevLett.119.050602}, trapped ions \cite{doi:10.1126/science.aad6320,Maslennikov2019,VanHorne2020}, optomechanical oscillators \cite{PhysRevX.7.031044,Elouard_2015,Brunelli_2015,PhysRevLett.112.150602,PhysRevLett.112.076803}, quantum dot \cite{Kennes_2013,S_nchez_2013,Sothmann_2014,Josefsson2018}, spins \cite{PhysRevLett.122.110601,PhysRevLett.123.240601,Bouton2021,PhysRevLett.125.166802,PhysRevLett.123.080602,PhysRevLett.128.090602,Zhao2017}, and superconducting circuits \cite{PhysRevApplied.17.064022,Ronzani2018,doi:10.1146/annurev-conmatphys-033117-054120}, etc. 
As suggested by recent works \cite{Jaramillo_2016,PhysRevLett.106.070401,PhysRevLett.120.100601,PhysRevLett.120.100601,PhysRevE.92.012110,Campisi2016,PhysRevE.94.052122,PhysRevE.98.052124,PhysRevE.98.052147,PhysRevLett.124.110606,PhysRevResearch.2.043247,Chen2019,PhysRevE.96.022143,Fogarty_2020,PhysRevE.96.030102,PhysRevLett.120.190602}, 
the performance of a QHE may be affected substantially by the criticality \cite{Campisi2016,PhysRevE.94.052122,PhysRevE.98.052124,PhysRevE.98.052147,PhysRevLett.114.050601,PhysRevLett.124.110606,PhysRevResearch.2.043247,Chen2019,PhysRevE.96.022143,Fogarty_2020,PhysRevE.96.030102,PhysRevLett.120.190602,PURKAIT2022128180}. %PHYSICAL REVIEW RESEARCH 2, 043247 (2020) 
Some of these studies suggest that the criticality might provide an advantage for improving the performance of QHEs.
For example, 
when modelling the working substance as a Lipkin-Meshkov-Glick, it is possible to achieve the Carnot efficiency of a quantum Stirling cycle in the low-temperature limit \cite{PhysRevE.96.022143}.
In Ref. \cite{Campisi2016}, it has been proved that the criticality can enable quantum Otto engines that approach the Carnot efficiency without sacrificing power.
Another study \cite{Chen2019}, proposed a thermodynamic cycle with two interaction-driven stokes, utilizing a 1-dimensional ultracold gas as the WS. it was found that the average work per particle approaches a maximum at the critical point.

However, it is not yet fully understood how universality at the critical point impacts the efficiency of QHEs. 
Resolving this issue requires determining the asymptotic behavior of a quantum heat engine as its parameters approach the critical point.
To this end, it would be beneficial to consider a model that has an analytical solution and exhibits a phase transition that can be easily observed through experiments. Recent research has demonstrated that a quantum phase transition (QPT) can occur in a system of only two constituents: a two-level atom and a bosonic mode \cite{PhysRevLett.115.180404,PhysRevLett.119.220601}. This system, described by the quantum Rabi model (QRM) and with an analytical solution, has been experimentally observed to exhibit QPT using trapped ions in a Paul trap \cite{Cai2021}.

In this paper, we investigate the critical behaviour of Stirling engine efficiency, based on WSs modeled as the QRM. 
Firstly, we analyze whether criticality is beneficial for improving the efficiency of such QHE by using the analytical solution of the QRM.
Furthermore, we derive the asymptotic behaviour of efficiency as a control parameter approaches the critical point, which illustrates a dependence on the critical exponent.
Additionally, We present numerical verifications that support our findings.
This result considerably improves our understanding of HEs utilizing criticality. 
Furthermore, we observe an extension of prior knowledge, where a Stirling cycle can approach the Carnot efficiency at the critical point, without the need for the low-temperature or high-temperature limit.

This paper is organized as follows. In Sec. \ref{sec:qhe}, the quantum Stirling heat engine and the QRM are introduced. In Sec. \ref{sec:critical_he}, we investigate the impact of criticality on the efficiency of a quantum Stirling HE, including the discussion the asymptotic behaviour of the efficiency in the vicinity of the critical point and numerical results. 
Sec. \ref{sec:conclusion} summarizes the findings. 

%%%%%%%%%%%%%%%%%%%%%%%%%%%%%%%%%%%%%%%%%
\section{\label{sec:qhe}The quantum heat engine}
%%%%%%%%%%%%%%%%%%%%%%%%%%%%%%%%%%%%%%%%%
Studies on QPTs usually focus on many-body systems in the thermodynamic limit, where the number of particles approaches infinity \cite{sachdev2011}.
However, recent discoveries have shown that a QPT can also take place in a small system consisting of only two constituents - 
a two-level atom and a bosonic mode, which is described by the QRM \cite{PhysRevA.85.043821,PhysRevB.69.113203,PhysRevA.70.022303,PhysRevA.81.042311,PhysRevA.82.025802,PhysRevA.87.013826,PhysRevLett.115.180404,PhysRevLett.119.220601}, it is one of the simplest models of light-matter interactions.
The QRM Hamiltonian can be expressed as (for simplicity, we set $\hbar =1$ here and after)
\begin{align}
H_{\text{Rabi}} & =\omega _{0} a^{\dagger } a+\frac{\Omega }{2} \sigma _{z} -\lambda \left( a+a^{\dagger }\right) \sigma _{x} ,
\end{align}
where $\displaystyle \sigma _{x,z}$ are Pauli matrices for a two-level system and $\displaystyle a$ ($\displaystyle a^{\dagger }$) is an annihilation (creation) operator for a cavity field. The cavity field frequency is $\displaystyle \omega _{0}$, the transition frequency $\displaystyle \Omega $, and the coupling strength $\displaystyle \lambda $.

In the $\Omega /\omega _{0}\rightarrow \infty $ limit, the low-energy effective Hamiltonian has been obtained in \cite{PhysRevLett.115.180404}.
In the normal phase, where the control parameter $g=2\lambda/\sqrt{\omega_{0}\Omega}<1$,
the effective Hamiltonian can be expressed as
\begin{align}
H_{\rm{np}}  =\omega_{0} a^{\dagger } a-\frac{\omega _{0} g^{2}}{4}\left( a+a^{\dagger }\right)^{2} -\frac{\Omega }{2},
\label{eq:h_normal}
\end{align}
with the qubit being in its ground state.
On the other hand, when $g>1$, the system is in the superradiant phase, and the effective Hamiltonian reads
\begin{align}
H_{\rm{sp}} & =\omega _{0} a^{\dagger } a-\frac{\omega _{0}}{4g^{4}}\left(a+a^{\dagger }\right)^{2} -\frac{\Omega }{4}\left( g^{2} +g^{-2}\right),
\label{eq:h_super}
\end{align}
it is in a displaced frame of the bosonic mode, the qubit's ground state now rotated toward the x-axis due to its strong coupling to the bosonic mode.
Eq. \eqref{eq:h_normal} can be diagonalized into $H_{\rm{np}} =\varepsilon_{\rm{np}} b^{\dagger} b-\Omega /2$, with the excitation energy $\varepsilon_{\rm{np}}  =\omega _{0}\sqrt{1-g^{2}}$,
which is real only for $g\leq 1$ and vanishes at $g=g_{C}=1$, locating at the QPT.
Similarly, Eq. \eqref{eq:h_super}, can be diagonalized into a similar form, with the ground state energy and the excitation energy to be replaced with $-(\Omega/4)\cdot(g^{2}+g^{-2})$ and $\varepsilon _{\rm{sp}} =\omega _{0}\sqrt{1-g^{-4}}$ (which is real for $g>1$), respectively.
Accordingly, for a WS modelled as the QRM, the $k$th eigenenergy can be expressed as: $E_{k}=k\varepsilon +E_{0}$, where $E_{0}$ is the ground state energy, and $\varepsilon$ is the excitation energy,
the partition function of such WS reads $Z =e^{-\beta E_{0}}/(1-e^{-\beta\varepsilon})$.
Applying the partition function, one can get the corresponding internal energy and the entropy:
\begin{align}
    U&=-\frac{\partial \ln Z}{\partial \beta}=E_{0} +\varepsilon \frac{e^{-\beta \varepsilon }}{1-e^{-\beta \varepsilon}}, \label{eq:t1}\\
    S&=\ln Z+\beta U=\frac{\beta \varepsilon e^{-\beta \varepsilon }}{1-e^{-\beta \varepsilon }}-\ln(1-e^{-\beta\varepsilon}).
    \label{eq:t2}
\end{align}

\begin{figure}
    \centering
    \includegraphics[width=0.45\textwidth]{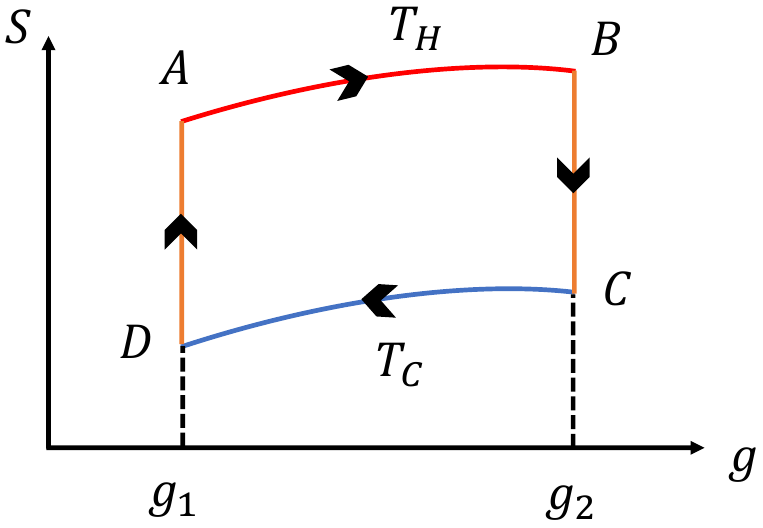}
    \caption{(a) Entropy-coupling diagram ($S-g$) of the Stirling cycle (clockwise or anticlockwise, depended on the work output), which consists of two isothermal processes and two isochoric processes. 
    Here $g$ is the only tunable parameter in completing the Stirling cycle, $g_{1}$ and $g_{2}$ are the corresponding parameters in the two isochoric processes $D\rightarrow A$ (or $A\rightarrow D$) and $B\rightarrow C$ (or $C\rightarrow B$) respectively.
    $T_{H}$ and $T_{C}$ are the corresponding temperatures of hot ($A\rightarrow B$ or $B \rightarrow A$) and cold ($C\rightarrow D$ or $D\rightarrow C$) isothermal processes, respectively.}
    \label{fig:cycle}
\end{figure}

In the following of this section, we will build a quantum Stirling heat engine, which works through performing a series of Stirling cycle. 
As depicted in Fig.\ref{fig:cycle}, the Stirling cycle consists of four thermodynamic processes acting on the working substance, including two isothermal process:
$A\rightarrow B$ and $C\rightarrow D$, 
and two isochoric processes:
$D\rightarrow A$ and $B \rightarrow C$.
The capital letters `A, B, C, D' represent four Gibbs states. 
During isochoric processes, the system is in equilibrium with a hot (cold) reservoir at an inverse temperature of $\beta_{H}=1/(k_{B}T_{H})$ ($\beta_{C}=1/(k_{B}T_{C})$), with $T_{H(C)}$ represents the temperature of the hot (cold) reservoir, the Hamiltonian in this process remains constant.

The efficiency of a thermodynamic cycle is determined by
$\eta=W/Q_{\text{in}}$,
where
$W=Q_{\text{DA}}+Q_{\text{AB}}+Q_{\text{BC}}+Q_{\text{CD}}$ is the output work of one cycle, 
with $Q_{XY}$ representing the heat transfer during the process $X\rightarrow Y$.
$Q_{\text{in}}=Q_{\text{DA}}+Q_{\text{AB}}$ represents the input heat of the Stirling cycle. 
Note that in the isothermal processes, $Q_{AB}=T_{H}(S_{B}-S_{A})$ and $Q_{CD}=T_{C}(S_{D}-S_{C})$;
while in the isochoric processes, $Q_{DA}=U_{A}-U_{D}$ and $Q_{BC}=U_{C}-U_{B}$. 
Accordingly, the efficiency can be written as 
\begin{align}
    \eta = \frac{\eta_{C}+\Sigma_{1}+\Sigma_{2}}{1+\Sigma_{2}}, \label{eq:eff_sim}
\end{align}
where 
\begin{align}
\Sigma_{1}&=\frac{T_{C}}{T_{H}}\cdot\frac{\Delta S_{AD}-\Delta S_{BC}}{\Delta S_{AB}}+\frac{Q_{BC}}{Q_{AB}},  \\
\Sigma_{2}&=\frac{Q_{DA}}{Q_{AB}},
\end{align}
with $\Delta S_{XY}\equiv S_{Y}-S_{X}$.
What needs to be emphasized is that the derivation of Eq. \eqref{eq:eff_sim} is independent of the WS.

Be reminded of the celebrated Carnot heat engine, which operates on the Carnot cycle, it consists of two driven isothermal processes and two adiabatic processes, which efficiency reads $\eta_{C}=1-T_{C}/T_{H}=\Delta T/T_{H}$,
with $\Delta T\equiv T_{H}-T_{C}$.
It provides an upper bound on the efficiency of any classical thermodynamic heat engine.
If the two adiabatic processes of the Carnot cycle are replaced by two isochoric processes, 
we get the Stirling cycle, which efficiency is given in Eq. \eqref{eq:eff_sim}.
The difference between Eq. \eqref{eq:eff_sim} and the Carnot bound comes from $\Sigma_{1}$ and $\Sigma_{2}$, 
from a more intuitive prospective, it results from the fact that the isochoric processes are irreversible.
In the following section, we will demonstrate that, by exploiting the criticality of the QRM, 
the difference between the Stirling cycle efficiency given in Eq. \eqref{eq:eff_sim} and the Carnot cycle efficiency, can be eliminated asymptotically.

%%%%%%%%%%%%%%%%%%%%%%%%%%%%%%%%%%%%%%%%%
\section{\label{sec:critical_he}The critical behavior of quantum Stirling engine with WS modelled as the QRM}
%%%%%%%%%%%%%%%%%%%%%%%%%%%%%%%%%%%%%%%%%
In this section, 
we consider a Stirling cycle with WS modelled as the QRM,
we assume that it operates in the normal phase, and $g$ is the only tunable parameter of the Hamiltonian in completing the Stirling cycle,
therefore, the parameter $g$, together with the inverse temperature $\beta$, determine the Gibbs states. 
We will show that, when the thermodynamic cycle introduced in the above section satisfies the following two conditions: 
1) $g_{C}-g_{1}$ is finite; 2) $g_{C}-g_{2}\rightarrow 0$,
the corresponding efficiency will approach the Carnot limit. 

Firstly, let us look at the heat transfer $Q_{DA}$, which can be rewritten as $Q_{DA}=\int_{T_{C}}^{T_{H}}[dU(\beta,g_{1})/dT]dT$, the integrand can be expressed as
\begin{align}
    \frac{dU(\beta,g_{1})}{dT}= C\big[\beta\varepsilon(g_{1})/2\big], \label{eq:dudt}
\end{align}
we set the Boltzmann constant $k_{B}=1$ here and after, the explicit expression of the heat capacity reads $C(x)=x^{2}\mathrm{Csch}^{2} (x)$.
For $x\in (0,\infty)$, the function $C(x)$ decreases monotonically with $x$, furthermore, $\lim_{x\rightarrow 0}C(x)\rightarrow 1$ and $\lim_{x\rightarrow \infty}C(x)\rightarrow 0$.
According to these properties of $C(x)$, we have the following inequality,
\begin{align}
    \Big|\frac{Q_{DA}}{Q_{AB}}\Big|< \eta_{C}\cdot\frac{C[\frac{\beta_{H}\varepsilon(g_{1})}{2}]}{\Delta S_{AB}}, 
    \label{eq:bound_daab}
\end{align}
this inequality gives an upper bound for the ratio $|Q_{DA}/Q_{AB}|$.
Secondly, we will derive an upper bound for  $|\Sigma_{1}-Q_{BC}/Q_{AB}|=  
|-(T_{C}/T_{H})\cdot\Delta S_{BC}/\Delta S_{AB}+Q_{BC}/Q_{AB}|$, and discuss under what conditions, this term will vanish.
Let's begin with looking at the derivation of the entropy $S$ with respect to the temperature $T$, which can be written as 
\begin{align}
\frac{dS(\beta,g)}{dT}=\frac{1}{T}\cdot C\Big[\frac{\beta\varepsilon(g)}{2}\Big], \label{eq:dsdt} 
\end{align}
according to \eqref{eq:dudt} and \eqref{eq:dsdt}, and by using the properties of $S(x)$ and $C(x)$,  the following relation can be obtained 
\begin{align}
    &\Big|-\frac{T_{C}}{T_{H}}\frac{\Delta S_{BC}}{\Delta S_{AB}}+\frac{Q_{BC}}{Q_{AB}}\Big| < \frac{2\eta_{C}}{\Delta S_{AB}} \cdot C\left[\frac{\beta_{H}\varepsilon(g_{2})}{2}\right] , \label{eq:bound_term1}
\end{align}
likewise, we get the following relation
\begin{align}
   \Big|\frac{T_{C}}{T_{H}}\frac{\Delta S_{AD}}{\Delta S_{AB}}\Big| < \frac{\eta_{C}}{\Delta S_{AB}}\cdot C\left(\frac{\beta_{H}\varepsilon(g_{1})}{2}\right). 
   \label{eq:bound_term2}
\end{align}
Combing with the expressions of $\Sigma_{1}$ and $\Sigma_{2}$, it is not hard to see that, when the upper bounds given in inequalities \eqref{eq:bound_daab}, \eqref{eq:bound_term1} and \eqref{eq:bound_term2} tends to zero,
then $\Sigma_{1}$ and $\Sigma_{2}$ will be eliminated.
There are two ways to achieve this goal: 1) let $\eta_{C}=\Delta T/T_{H} = 0$, and $\Delta S_{AB}$ is finite; 2) let $\Delta S_{AB}\rightarrow \infty$. 
Usually, the efficiency of a thermodynamic cycle is expected to be as high as possible, therefore, the second way is preferable. 
Based on Eq. \eqref{eq:t2} and the fact that $\varepsilon_{\rm{np/sp}}(g_{C})=0$, it can be inferred that the degeneracy and the entropy of the QRM at the critical point are infinite. 
Thus, when $g_{C}-g_{1}$ is finite and $g_{C}-g_{2}\rightarrow 0$, 
$S_{A}$ is finite and $S_{B}\rightarrow \infty$, 
as a result, $\Delta S_{AB}\rightarrow \infty$. 
It is noteworthy that, since $C(\beta\varepsilon/2)$ decreases monotonically with increasing $T=1/\beta$, and $\lim_{T\to 0}C(\beta\varepsilon/2)\to 0$, 
it follows that, when $g_2$ approaches the critical point from the normal phase with $Q_{AB}$ and $\Delta S_{AB}$ tending towards divergence, or $\beta$ increasing (i.e., temperature decrease, leading to a decrease in $C(\beta\varepsilon/2)$), both cases result in the cycle's efficiency trending towards the Carnot efficiency.
In addition, we consider cases where $g_2$ crosses the critical point to enter the super-radiant phase. While the isothermal process of crossing the critical point is unachievable in practice, we assume an ideal Stirling cycle for our analysis. In the super-radiant phase, the ground state has a finite degeneracy of two, giving rise to finite values of $\Delta S_{AB}$ and $Q_{AB}=T_H \Delta S_{AB}$. By incorporating equations  inequalities \eqref{eq:bound_daab}, \eqref{eq:bound_term1} and \eqref{eq:bound_term2}, as well as the expressions for $\Sigma_1$ and $\Sigma_2$, we can reach the following conclusion: for cases where $g_{2}>g_{C}$, the degree to which efficiency can approach the Carnot efficiency is dependent on the temperature of heat reservoirs, only at low-temperatures limit, where $C(\beta\varepsilon/2)$ approaches zero, the efficiency tends towards the Carnot efficiency.

We may conclude that, for a Stirling cycle with WS modelled as the QRM, $g_{C}-g_{1}$ is a positive finite value and $g_{C}-g_{2}\rightarrow 0^{+}$ are sufficient conditions for the efficiency to approach the Carnot bound.
This is an interesting result: the sufficient conditions for the Stirling cycle efficiency to approach the Carnot efficiency does not imply the vanishing of entropy production, since the two isochoric processes are not reversible. 
Additionally, for cases where $g_{2}>g_{C}$, $g_{C}-g_{1}$ is a positive finite value and $T_{H}\to 0$ are sufficient conditions for the efficiency tends towards the Carnot efficiency.

Furthermore, we will discuss the critical behaviour of the efficiency in this paragraph. 
When $g_{2}$ is sufficiently close to $g_{C}$, with $\Sigma_{i}\ll 1$ for $i\in {1,2}$, then to the first order in $\Sigma_{1}$ or $\Sigma_{2}$, Eq. \eqref{eq:eff_sim} can be approximated as $\eta-\eta_{C} \approx \Sigma_{1}+(1-\eta_{C})\Sigma_{2}$. By defining $\alpha (g_{2}) =[T_{C}(\Delta S_{AD} -\Delta S_{BC})+Q_{BC}+(T_{C}/T_{H})Q_{DA}]/T_{H}$, 
the approximated relation can be rewritten as $\eta-\eta_{C} \approx -\alpha(g_{2})/\Delta S_{AB}$.
On the other hand, it is not difficult to verify that the entropy change $\Delta S_{AB}$ in the limit $g_{2}\rightarrow g_{C}$ is given by $\lim_{g_{2}\rightarrow g_{C}} \Delta S_{AB} \rightarrow -\ln \varepsilon(g)$. Combing this with the fact that the excitation energy in both phases near the critical point of the QRM, $\varepsilon_{\rm{np}}$ and $\varepsilon_{\rm{sp}}$, vanishes as $\varepsilon(g)\propto |g_{C}-g|^{z\nu}$, where $\nu$ ($z$) is the (dynamical) critical exponent \cite{PhysRevLett.115.180404}, we have the following relation 
\begin{align}
    \lim_{|g_C-g_2|\to 0}\eta_{C}-\eta \to \frac{\alpha (g_{2})}{\ln \left(g_{C}-g_{2}\right)^{z\nu}}. \label{eq:eff_scale}
\end{align}
Eq. \eqref{eq:eff_scale} is valid for any WSs with homogeneous energy-level-spacing $\varepsilon(g)\propto |g_{C}-g|^{z\nu}$ in the vicinity of the critical point, and undergoes thermodynamic cycle 
depicted in Sec. \ref{sec:qhe}.
From Eq. \eqref{eq:eff_scale}, we readily know that the efficiency $\eta$ approaches the Carnot efficiency $\eta_c$ when the  $g_2\to g_c$. It means that, by approaching the critical coupling point, the efficiency of the heat engine can be greatly improved, approaching the Carnot efficiency. Furthermore, the key equation \eqref{eq:eff_scale} describes the asymptotic behavior of a quantum heat engine as its parameters approach the critical point, and illustrates how the efficiency depends on the critical exponent. This asymptotic behavior is characterized by a logarithmic divergence in the denominator of  Eq. \eqref{eq:eff_scale}.

%%%%%%%%%%%%%%%%%%%%%%%%%%%%%%%%%%%
\begin{figure}[!h]
    \centering
    \includegraphics[width=0.45\textwidth]{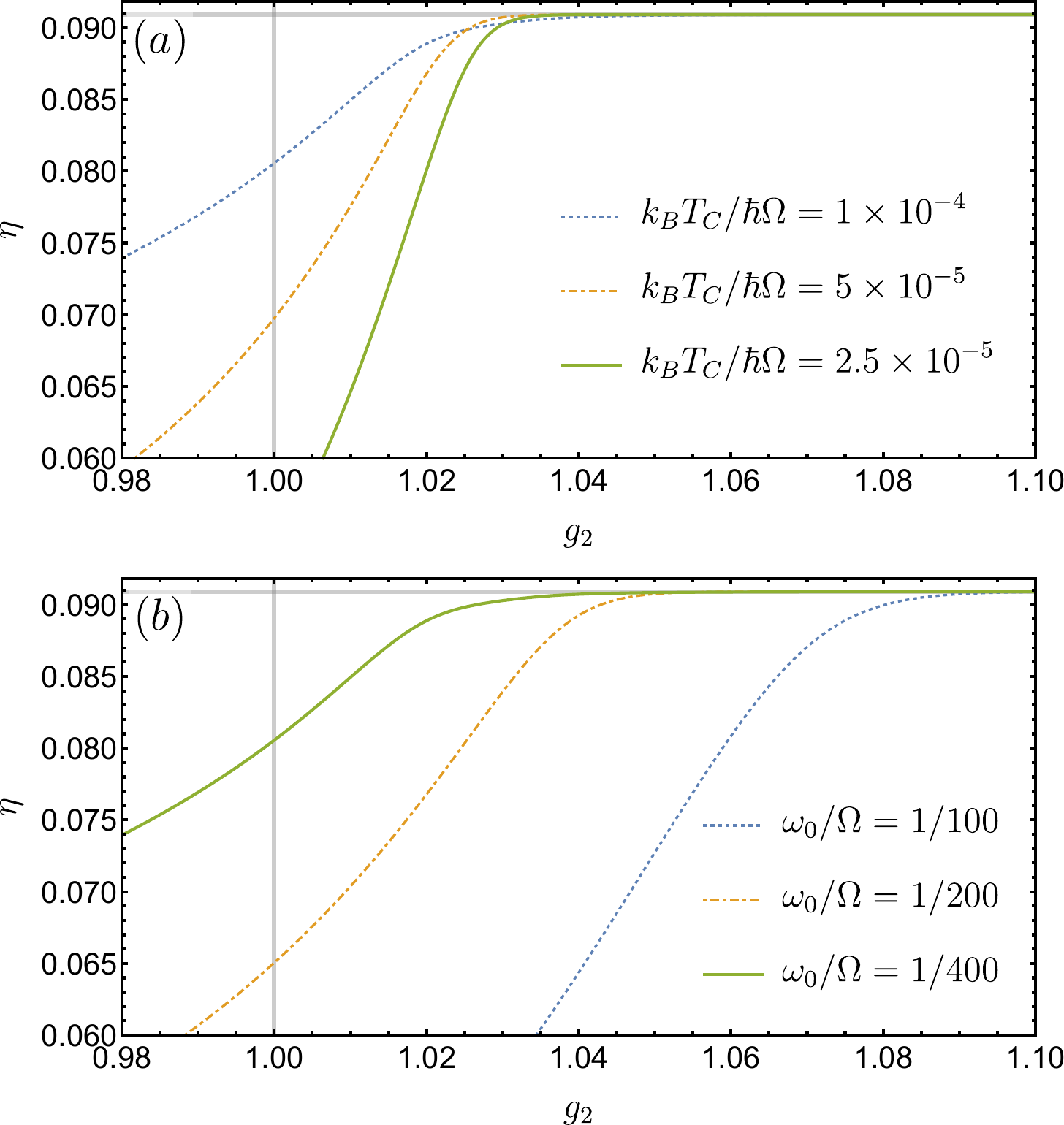}
    \caption{Numerical evidence for critical Stirling engine of approaching Carnot efficiency.  (a) The efficiency versus $g_{2}$ for different temperature, we set $\omega_{0}/\Omega=1/400$. 
    (b) The efficiency versus $g_{2}$ for different $\omega_{0}/\Omega$, the thermodynamics limit is achieved when $\omega_{0}/\Omega\rightarrow 0$, we set
    $k_{B}T_{C}/\hbar\Omega=1\times 10^{-4}$.
    %$k_{B}T_{C}/\hbar\Omega=2\times 10^{-5}$.  
    The vertical gray solid line represents the critical point $g=1$, while the horizontal gray solid line represents the Carnot efficiency.
    Other parameters: $\Omega=5$ GHz,
    $T_{H}=T_{C}+\Delta T$ with $\Delta T=T_{C}/10$.}
    \label{fig:eff_1}
\end{figure}
%%%%%%%%%%%%%%%%%%%%%%%%%%%%%%%%%%%%%%%%%
\section{\label{sec:numerical}Numerical results}
%%%%%%%%%%%%%%%%%%%%%%%%%%%%%%%%%%%%%%%%%
\begin{figure}[!h]
    \centering
    \includegraphics[width=0.45\textwidth]{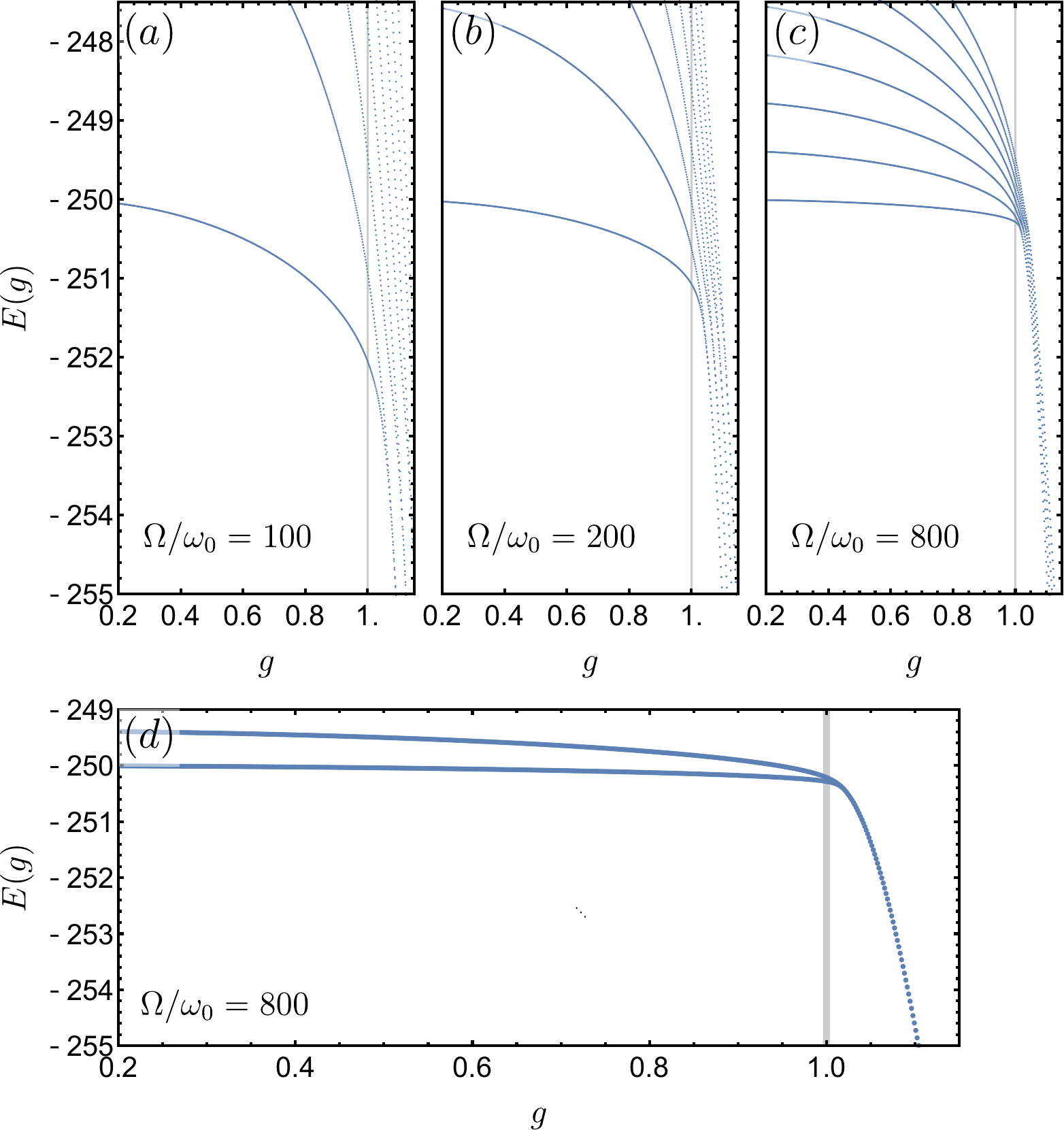}
    \caption{The variation of the QRM degeneracy near the critical point. The upper panel: the lowest eight energies of the QRM versus $g$ with different transition frequencies, (a) $\Omega/\omega_{0}=100$, (b) $\Omega/\omega_{0}=200$, (c) $\Omega/\omega_{0}=800$.
    The lower pannel (d): the lowest two energies of the QRM versus $g$ for $\Omega/\omega_{0}=800$.
    The vertical gray solid line locates the critical point in the thermodynamic limit.
    }
    \label{fig:ener_spec}
\end{figure}

In this section, numerical demonstrations are presented to support the conclusions drawn in the previous section. 
In deriving the numerical results, we first perform numerical diagonalization on the Hamiltonian of the QRM, and subsequently calculate thermodynamic quantities based on the obtained results.

Fig. \ref{fig:eff_1} (a) displays the relationship between the efficiency and $g_{2}$ at different temperatures. 
As $g_{2}$ increases, the efficiency seems to converge to the Carnot efficiency (depicted as a horizontal gray solid line). 
Due to the finite size effect, the convergence point $g_{2} = g_{m}$ is greater than $g_{C}$, represented by the vertical gray solid line. 
Efficiency converging to the Carnot efficiency at $g_{2}=g_{m}$ indicates that for any $g_{2}\geq g_{m}$, $\eta=\eta_{C}$. 
However, although in the previous section we discussed that $\eta_{C}-\eta\to 0$ as $g_{C}-g_{2}\to 0^{+}$, it is only under low-temperature limit that the efficiency can approach the Carnot efficiency when $g_{2}>g_{C}$. 
The efficiency in Fig. \ref{fig:eff_1} appearing to approach the Carnot efficiency at $g_{2}>g_{C}$ is attributed to the low temperatures considered, which resulted in an efficiency close to the Carnot efficiency.
At a finite temperature, $g_{m}$ represents the maximum efficiency point. Once $g_{2}$ exceeds this point and enters the super-radiant phase, the efficiency starts decreasing. 
In the thermodynamic limit, $g_{m}=g_{C}$, resulting in the efficiency achieving the Carnot efficiency.
In addition, when the temperature drops, the efficiency change near $g_{m}$ in the normal phase is more rapid. 
This observation can be explained by our previous discussion:
to achieve a constant difference $\eta_{C}-\eta$, lower temperatures (i.e., higher $\beta$ values) can relax the proximity requirements of $g_{2}$ to the critical point and therefore provide earlier observations of the convergence.
In Fig. \ref{fig:eff_1} (b), we present how the efficiency change with $g_{2}$ for different values of $\omega_{0}/\Omega$, it is worth noting that when $\omega_{0}/\Omega\rightarrow 0$ (or equivalently, as $\Omega/\omega_{0}\to\infty$) the QRM moves closer to attaining the thermodynamic limit. 
It is apparent that, $g_{m}$ will approach the critical point of the QRM as $\omega_{0}/\Omega$ tends towards the thermodynamics limit. 
Consequently, our numerical findings suggest that, in the thermodynamic limit, the efficiency of the thermodynamic cycle illustrated in Fig. \ref{fig:cycle}, which employs the QRM as working substance, will reach the Carnot efficiency when $g_{C}-g_{2}\to 0^{+}$. 

To better comprehend the aforementioned discussion and further elucidate the finite size effect, we display the lowest several energy levels of the QRM with finite size in Figure \ref{fig:ener_spec}.
The top panel depicts how the first eight lowest energies evolve as $g$ varies for different $\Omega/\omega_{0}$ values.
It can be inferred from the figure that, as $\Omega/\omega_{0}$ increases, the energy level spacing near $g=g_{C}$ decreases.
Additionally, at the critical point (depicted as a vertical gray solid line), the energy levels tends to become highly degenerate, beyond the critical point, the degeneracy on the right side of the critical point is higher than the left side.
The bottom panel illustrates the first two energy levels of the QRM when $\Omega/\omega_{0}=800$.
As is evident from the figure, on the right side of the critical point, the ground state is degenerate.
As such, in the low-temperature limit, if $0<(g_{C}-g_{1})/g_{C}\ll 1$ (i.e., $g_{1}$ is in the normal phase and is far away from the critical point) and $g_{2}>g_{C}$ while the ground state corresponding to $g_{2}$ is degenerate, then the efficiency of the Stirling cycle depicted in Figure \ref{fig:cycle} will approach the Carnot efficiency.
%%%%%%%%%%%%%%%%%%%%%%%%%%%%%%%%%%%%%%%%%%%%%
\section{\label{sec:conclusion}Conclusion and outlook}
%%%%%%%%%%%%%%%%%%%%%%%%%%%%%%%%%%%%%%%%%%%%%
In this paper, we explore the influence of quantum criticality on the efficiency of a Stirling cycle that utilize the QRM model as their WS. 
We assume that the effective coupling constant $g$ is the only tunable parameter of the Hamiltonian needed to complete the thermodynamic cycle.
The Stirling cycle comprises two isochoric processes with corresponding coupling constants of $g_{1}$ and $g_{2}$ such that $g_{1}<g_{2}$.
Our results demonstrate that the efficiency approaches the Carnot efficiency when the thermodynamic cycle satisfies the following conditions:
1) $g_{C}-g_{1}$ is positive and finite, and 2) $g_{C}-g_{2}\to 0^{+}$.
Furthermore, we derive an analytical expression for the efficiency of the quantum Stirling engine when $g_{2}$ is in the vicinity of the critical point.
Our analysis reveals that as $g_{2}$ approaches the critical point, the asymptotic behavior of a quantum heat engine is characterized by a logarithmic divergence in the denominator. 
Additionally, we provide a numerical demonstration of our analytical findings, which includes an explicit analysis of the finite size effect.
This study deepens our understanding of how criticality affects the performance of a Stirling heat engine, while also advancing our appreciation of criticality.

\begin{acknowledgments}
This work was supported by the Natural Science Foundation of Guangdong Province (Grant No.2017B030308003), 
the Key R\&D Program of Guangdong province (Grant No. 2018B030326001), 
the Science, Technology and Innovation Commission of Shenzhen Municipality (Grant No.JCYJ20170412152620376 and No.JCYJ20170817105046702 and No.KYTDPT20181011104202253), 
the National Natural Science Foundation of China (Grant No. 12135003, No. 12075025, No.12047501, No.11875160, No.U1801661 and No.11905100),
the Economy, Trade and Information Commission of Shenzhen Municipality (Grant No.201901161512), 
the Guangdong Provincial Key Laboratory(Grant No.2019B121203002).
\end{acknowledgments}

%%%%%%%%%%%%%%%%%%%%%%%%%%%%%%%%%%%%%%%%%%%%%%%%%%%%%%%%%%%%%%%%%%%%%%

\bibliography{manuscript}% Produces the bibliography via BibTeX.

\end{document}